# Electron beam modification of vanadium dioxide oscillators

**Maksim Belyaev** [*,1], **Andrey Velichko**[1], **Vadim Putrolaynen**[1], **Valentin Perminov**[1], **and Alexander Pergament**[1]

[1] Petrozavodsk State University, Petrozavodsk, 185910, Russia

The paper presents the results of a study of electron-beam modification (EBM) of $VO_2$-switch *I-V* curve threshold parameters and the self-oscillation frequency of a circuit containing such a switching device. EBM in vacuum is reversible and the parameters are restored when exposed to air at pressure of 150 Pa. At EBM with a dose of 3 C/cm$^2$, the voltages of switching-on ($V_{th}$) and off ($V_h$), as well as the OFF-state resistance $R_{off}$, decrease down to 50% of the initial values, and the oscillation frequency increases by 30% at a dose of 0.7 C/cm$^2$. Features of physics of EBM of an oscillator are outlined considering the contribution of the metal and semiconductor phases of the switching channel. Controlled modification allows EBM forming of switches with preset parameters. Also, it might be used in artificial oscillatory neural networks for pattern recognition based on frequency shift keying.

**1 Introduction** Oxides and oxide interfaces are becoming increasingly important in the major trends of development of contemporary electronics [1]. Specifically, vanadium dioxide, with its metal-insulator phase transition [2], can be used for engineering of diverse electronic devices [3].

Electrical oscillations in $VO_2$-based systems have been widely recognized since 2008 [4] and even some time earlier [5,6]; recent research on such systems application in oscillatory neural networks (ONN) [3,7] and band generators [8] have resurrected scientists' interest in this area.

Controllability of metal-insulator transition (MIT) parameters of $VO_2$ films is an important problem [2, 5], since it allows, e.g., tuning the switching threshold voltage, which could turn out to be necessary for a number of applications. Various types of modification methods have been reported: EBM [9, 10], ion irradiation [11, 12, 13], impurity doping [14], hydrogenation [15], electrostatic doping using an ionic liquid [16] and mechanical strain [2]. The effect of these modifications on MIT parameters is very similar regardless of different physical mechanisms of modification. First of all, a controlled decrease of the transition temperature $T_t$ occurs, also the value of conductivity jump and the hysteresis loop width may change. Distinct advantage of irradiation methods [9-13] lies in their capability to affect local micro- and nanoscale areas. In this case EBM damages crystal lattice to a lesser degree as compared with the ion irradiation technique thus allowing us to change MIT parameters under more strict control in the irradiation area and to achieve their subsequent recovery.

In our previous publication [9], we have investigated electron-beam modification and the process of subsequent recovery of MIT parameters of $VO_2$ films. It has been shown that the level of EBM, that is, the resistance change value, is by an order of magnitude higher in $VO_2$ metallic phase, i.e. above the transition temperature ($T_t$ ~ 340 K in bulk [2]), compared with that in the semiconducting phase. Also, the regularities of EBM influence on the electrical switching effect have been established, and, particularly, the threshold voltage of a switching device has been found to decrease under the action of e-beam irradiation [9]. In our opinion the modification mechanism could be associated with oxygen vacancies generation at the boundary with vacuum (because of oxygen atoms drifting into vacuum) and their diffusion into film depth due to concentration gradient [9]. Generation of donor-type oxygen vacancies leads in its turn to electrons concentration increase in the conduction band and to MIT temperature decrease by Mott mechanism [17]. Low MIT temperature results in the fact that EB-modified films can be in metallic state even at room temperature. The results of Raman-spectroscopy studies of EB-modified films [10] may verify this fact.

The phenomenon of self-sustained oscillations in a single oscillator circuit is not possible without a non-linear element with the S-shaped *I-V* characteristic, the physics of which may be explained by the model of critical temperature [18]. According to this model, Joule heating of a film by passing current up to MIT temperature $T_t$ results in conduc-tivity jump thus accounting for formation of areas with negative differential resistance (NDR) of S-type at *I-V* characteristic.

---

\* Corresponding author: e-mail biomax89@yandex.ru

During the course of oscillations, alternating metallic and semiconductor states of the switching channel are observed, therefore the question of the EBM impact on the oscillation frequency remains open. Such oscillators, being connected in a network, can serve as prototypes of ONN with a pattern recognition function [3, 7]. In this regard, the development of alternative methods of information input into the system, by modifying the threshold characteristics of its constituent elements, is also an important issue.

Thus, the aim of this work is to study the physics of EBM influence on the frequency of the single oscillator assembled on the basis of a $VO_2$ switching structure. In addition, we will determine conditions for the repeated recovery (after EBM) of the oscillator initial state, with a view to its further use for the ONN implementation.

**2 Experimental procedure** Thin films of vanadium dioxide were obtained by the magnetron sputtering method on r-cut sapphire substrates using an AJA Orion system. The films were prepared in two stages: first, an amorphous vanadium oxide layer was deposited at room temperature while sputtering of a metal target (at the argon and oxygen flow rates of 14 and 2 $cm^3$/min, resp., and pressure of 5 mTorr) followed by an annealing in pure oxygen atmosphere at pressure 10 mTorr and temperature 480 ºC for 50 min. The temperature dependences of resistivity of these films measured by means of the four-probe technique (Fig. 1a) demonstrated a MIT at about 68 ºC, with the ~10 ºC hysteresis loop width and ~$10^3$ resistivity jump, which was characteristic of vanadium dioxide.

Planar devices were formed by the optical lithography (Heidelberg Instruments µPG 101) and lift-off processes. The electrodes were two-layer V-Au metal contacts with the overall thickness of about 50 nm (30 nm of vanadium and 20 nm of gold). The electrode width was 10 µm, and the inter-electrode gap was 3 µm (see insert in Fig. 1a).

For electrical measurements, a two-channel sourcemeter Keythley 2636A was used. To observe self-oscillations, a single oscillatory circuit (Fig. 1b) was assembled consisting of the load resistor $R_s$ = 50 kΩ, series-connected switch, and a power supply with the output voltage $V_{DD}$ = 60V. A capacitance $C$ = 17 nF was connected to the switch in parallel, the nominal value of which significantly affected the frequency of oscillation [3]. To limit the maximum values of the capacitance discharging current through the switch, a resistor $R_i$ = 50 Ω was connected in series with it.

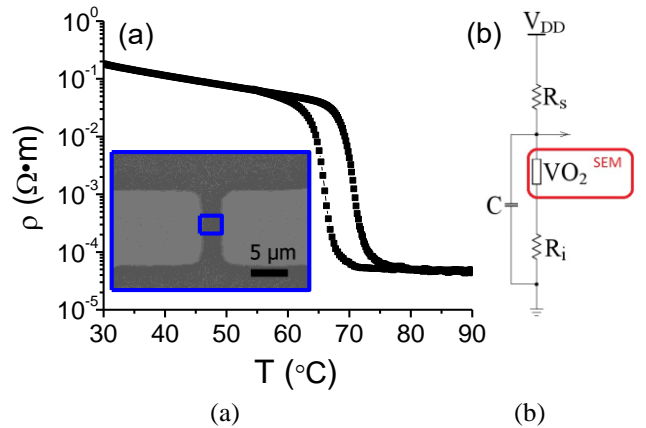

**Figure 1** Temperature dependence of resistance of the film (a) and the image of a structure under study (insert); oscillator circuitry (b)

To study the effect of EBM on the switching threshold characteristics, the structure was placed in the scanning electron microscope (SEM) Hitachi SU1510 sample chamber, whilst all the rest circuit was in air, under normal conditions, and it was shielded against external electromagnetic interference to minimize the noise affecting the circuit operation dynamics. The studies of oscillation dynamics were carried out with a four-channel oscilloscope Picoscope 5442B whose maximum sampling rate was 1 GS/sec.

Irradiation was carried out under high vacuum ($P < 1$ Pa), and the reverse recovery of the parameters was conducted at a higher pressure $P$ = 150 Pa. To recover the parameters air was let in the chamber at normal conditions. The e-beam current was measured by a Faraday cup, and it was ~7.8 nA at accelerating voltage of ~10 kV. The irradiated area was ~ 500 µ$m^2$ and it completely covered the switch operating space (outer frame in insert of Fig. 1a).

The measurements of the first harmonic oscillation frequency ($F_1$), as well as the threshold switching characteristics ($V_{th}$, $V_h$), were carried out in an automatic mode, on a sample of $10^4$ oscillation periods, using a specially developed software for Picoscope 5442B.

**3 Results and discussion** The *I-V* characteristic of the structure before EBM is shown in Fig. 2a (curve 1). It does not demonstrate S-type switching within the given voltage range, since the structure possesses a rather high resistance ~ 30 kΩ, and the switching threshold voltage reaches a value of $V_{th}$ ~ 30 V. At so high $V_{th}$, first electrical switching into the metallic state usually leads to irreversible degradation of the structure due to the strong heating of the film in the interelectrode gap.

As is shown in our study [9], EBM significantly changes the parameters of MIT, viz. it reduces the MIT temperature $T_t$ and results in an overall decrease in resistance of both the semiconductor and metal phases ($R_{off}$, $R_{on}$). Reduction of $R_{off}$ and $T_t$, in turn, leads to a decrease in the value of $V_{th}$ [9]. Therefore, to obtain stable and repeatable switching, within the used voltage range (up to 5 V), it is necessary to reduce the $V_{th}$, e.g., by means of EBM.

Irradiation is carried out capturing the entire switch operating space (outer frame in insert of Fig. 1a). Modification of the *I-V* curves of the structures during irradiation is presented in Fig. 2a. One can see a series of curves showing a decrease in the film resistance in the insulator phase, $R_{off}$. This leads to an increase in the current flowing across the structure, and consequently to an increase of the maximum temperature in the channel, when sweeping. At some moment during voltage sweep, the temperature in the channel reaches a critical value ($T_t = 68$ °C), and electrical switching into the conducting state occurs. The exposure dose value is ~ 0.5 C/cm$^2$ at this time. It should be noted that irradiation of the sample in a small local area (inner frame in insert of Fig. 1) forms a local conductive channel. In addition, this method of irradiation eliminates the self-formation of several channels (which would be visible as multiple separate electrical switching events in *I-V* curves) and reduces the threshold voltage variance. Furthermore, devices with channels of arbitrary shape and with any number of channels can be produced by such a local e-beam modification.

Thus, EBM of the channel region provides an *I-V* characteristic with repeatable S-shaped electrical switching and reduced $V_{th}$. This process, by analogy with electroforming [3], may be called as 'electron-beam forming' of the channel.

A further increase in the irradiation dose leads to a further reduction of $R_{off}$ and $V_{th}$. Figure 2b shows the *I-V* curve variation with changing the exposure dose from 0.5 C/cm$^2$ to 3 C/cm$^2$. One can see that a slight decrease in $V_h$ also occurs during irradiation, which might be accounted for by some reduction of the MIT reverse (metal-to-insulator) temperature. Such a behavior of the metal state during EBM has also been observed in [9] that apparently is connected with a shift of the MIT temperature of all the grains toward lower temperatures.

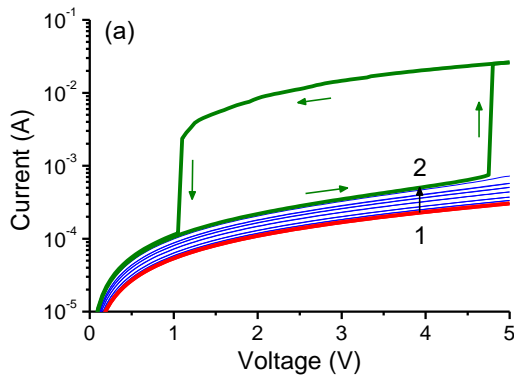

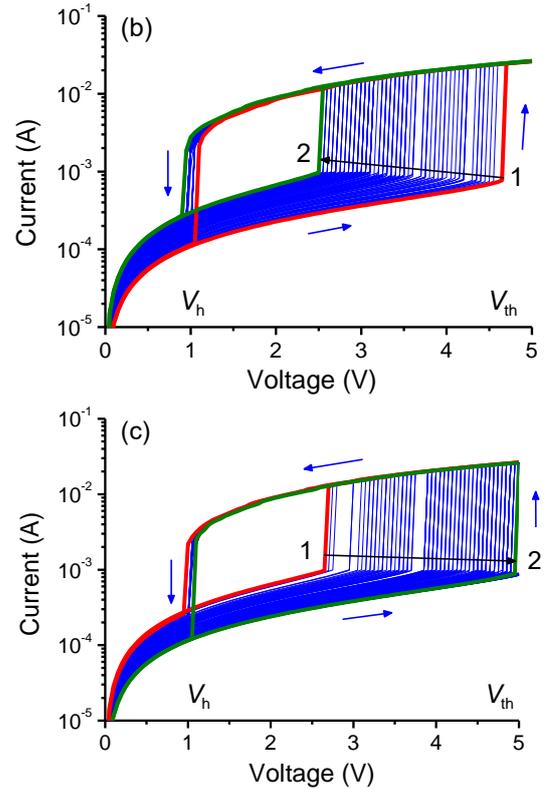

**Figure 2** Modification of *I-V* characteristic during the process of EBM forming (a): curve 1 – initial state; curve 2 – after irradiation with the exposure dose 0.5 C/cm$^2$. Modification of *I-V* characteristic at e-beam irradiation (b): curve 1 – initial *I-V* characteristic; curve 2 – after irradiation with the exposure dose 3 C/cm$^2$. Recovery process at air pressure 150 Pa (c): curve 1 – *I-V* characteristic before gas inlet; curve 2 – after recovery for 40 min.

To study the process of reverse recovery of the switch parameter values, the electron beam is turned off, and the pressure in the sample chamber is increased up to $P = 150$ Pa. Figure 2c shows a serial change of the structure *I-V* curve occurring during 40 minutes. In this case, an increase in the values of the switch parameters ($V_{th}$, $V_h$, $R_{on}$, $R_{off}$) toward their initial quantities (corresponding to the forming exposure dose of 0.5 C/cm$^2$) is observed. Some instability of recovery at the initial stage of the gas inlet is caused by a slight temperature change in the microscope chamber after a sharp increase of pressure. Thus, the air inlet results in the recovery of switching *I-V* characteristic parameters. After studying the electron beam exposure on the switch *I-V* characteristic, the circuit is converted into the mode of oscillation generation in order to identify the EBM impact on the dynamics of oscillations in a single circuit. The shape of voltage oscillogram was similar to that obtained in other studies [3, 4, 8, 19, 20]. Figure 3 shows the time dependences of the fundamental frequency $F_1$ of the oscillator, as well as the threshold ($V_{th}$) and holding ($V_h$) voltages, obtained from the voltage waveform, under alternation of the conditions of irradiation and recovery. Certain regions are allocated on the presented dependences, namely, those corresponding to the storage of the structure in vacuum (1), brief air inlet to 150 Pa (2), the process of evacuation (3), and the region of electron beam irradiation in vacuum (4).

It is seen that long aging of the structure in vacuum (1) does not change the main parameters, and the oscillation frequency stabilizes around some average value. The frequency scatter of about 2% is associated with a general instability of the threshold voltage described in [18].

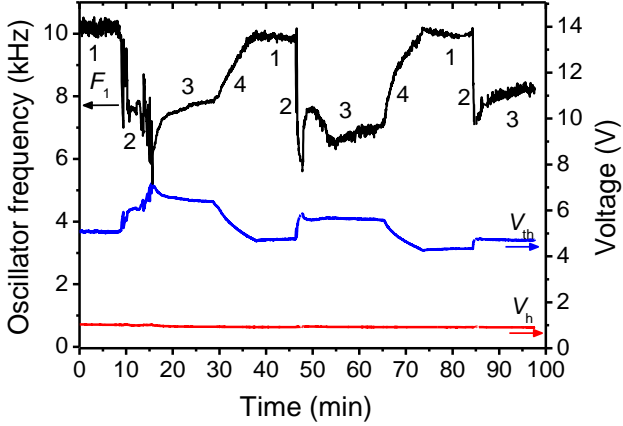

**Figure 3** Time diagrams of changes of the oscillation frequency $F_1$, threshold voltage $V_{th}$ and holding voltage $V_h$: 1 – storage in vacuum; 2 – air inlet to 150 Pa; 3 – vacuum pumping; 4 – electron beam exposure.

At the air inlet stage (2), the oscillation frequency drops sharply, which is caused by the raise of $V_{th}$, similarly to the recovery of the *I-V* curve parameters in Fig. 2c. Prolonged exposure to air can lead to the oscillation cease, which is due to the shift of the circuit operating point beyond the NDR region, since the value of $V_{th}$ increases. In this case, one has to select either an appropriate power supply voltage $V_{DD}$ or the value of series resistor. To prevent this, i.e. to keep the circuit in the self-oscillating mode permanently, the cycling of parameters is carried out in a certain relatively narrow range (Fig. 3).

At the subsequent vacuum pumping (3), the oxide film is partially reduced, which leads to a decrease in $V_{th}$ and a gradual increase in the oscillation frequency with a tendency to reach some stationary state. The electron-beam irradiation in vacuum (4) results in a significant increase of the frequency $F_1$ and decrease of $V_{th}$. Figure 4 shows the dependence of the parameters ($F_1$, $V_{th}$, $V_h$) on the specific exposure dose of electron irradiation $D$. One can see here an almost linear dependence of $V_{th}$ on the dose, whereas the dependence of frequency has a nonlinear form because of its complex dependence on $V_{th}$ and $R_{off}$. It is straightforward to understand that the oscillation period is composed of a capacitor charging time $t_s$ (through $R_s$) and the time of its discharge $t_m$ (through $R_{on}$), $T = t_s + t_m$. Using the circuit differential equation given in [3], the following estimates can be written:

$$t_s = \frac{CR_s}{x} \ln\left[\frac{V_{DD} - xV_h}{V_{DD} - xV_{th}}\right] \quad (1),$$

$$t_m = CR_{on} \ln\left[\frac{V_{th}}{V_h}\right] \quad (2),$$

whence

$$F_1 = C^{-1}\left[\frac{R_s}{x}\ln\left[\frac{V_{DD} - xV_h}{V_{DD} - xV_{th}}\right] + R_{on}\ln\left[\frac{V_{th}}{V_h}\right]\right]^{-1} \quad (3)$$

where $x=(R_s/R_{off}+1)$. Expression (3) is a more accurate estimate for $F_1$ than the formula obtained in our previous work [3].

As one can see from formula (3), the oscillation frequency $F_1$ depends on many parameters determined by the external circuit ($C$, $R_s$, $V_{DD}$) and VO$_2$-switch parameters ($V_{th}$, $V_h$, $R_{on}$, $R_{off}$). This accounts for the large spread of frequencies (ranging from kilohertz to megahertz) observed in different studies [3, 4, 8, 19, 20]. Obviously, EBM affects only the switch parameters.

Using formula (3) and the empirical relationships $V_{th}(D)$, $V_h(D)$, $R_{off}(D) = 1/(1.39 \cdot 10^{-4} + 5.71 \cdot 10^{-5} \cdot D)$ [Ω], $R_{on} \sim 250$ Ω, one can show (see curves 'Model' and 'Experiment' in Fig. 4) that the experimental and calculated frequencies have similar dependences on the dose.

When cycling 'air inlet/EBM', the threshold voltage $V_{th}$ corresponding to a certain $F_1$, as well as the $V_{th}$ change magnitude, gradually decline. This is apparently connected with the redistribution of the $V_{th}$ and $R_{off}$ contributions to the oscillation frequency. It should be noted that the holding voltage $V_h$, as can be seen from Figs. 3 and 4, is almost independent of an external impact, it is poorly responsive to EBM and oxidation and decreases only by 150 mV for three cycles. Such behavior can be explained by the fact that the structure resistance in the metal state $R_{on}$ is weakly dependent on the oxidation state, and the value of $V_h$ to a greater extent depends on the reverse phase transition temperature, which varies much more slowly than $R_{off}$.

It has been shown in our paper [9] that the intensities of EBM ($\xi_m$, $\xi_s$) and recovery times ($\tau_m$, $\tau_s$) of a VO$_2$ film depend on whether VO$_2$ is in the metallic or semiconducting phase. In the metallic phase, the characteristic time of recovery in air ($\tau_m \sim 100$ s) is lower than that in the semiconducting phase ($\tau_s \sim 10000$ s), that is $\tau_s/\tau_m \sim 100$, and the EBM intensity ratio is $\xi_m/\xi_s \sim 1.5$. During oscillation, the switching channel is periodically either in the S-phase (for $t = t_s$) or in the M-phase ($t_m$); that is why the oscillator EBM process has such a complex nature.

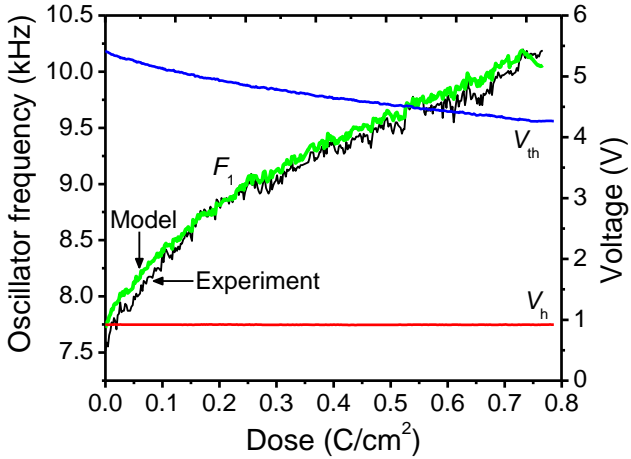

**Figure 4** Dependence of oscillation frequency $F_1$ (experiment and model), threshold voltage $V_{th}$ and holding voltage $V_h$ on dose at electron irradiation with accelerating voltage 10 kV and current density ~ 1.56 mA/cm$^2$.

We thus arrive at a conclusion that, when $t_s/t_m<1.5$, the processes of the metal phase modification would mostly contribute to EBM of an oscillator, while at $t_s/t_m>1.5$ the semiconductor phase contribution to EBM is more significant. During the process of recovery at $t_s/t_m<100$, on the contrary, the contribution of the metal phase is greater than that of the semiconductor one.

Experimentally, the time ratio ($t_s/t_m$) could be easily estimated from the switch voltage waveform. Theoretically, this ratio can be estimated, using Equations (1) and (2), from the formula:

$$\frac{t_s}{t_m} \sim \frac{R_{off}}{R_{on}} \ln\left[\frac{V_{DD} - xV_h}{V_{DD} - xV_{th}} - \frac{V_{th}}{V_h}\right] \quad (4)$$

Relation (4) does not depend on the capacitance $C$, and hence on the oscillation frequency, and it is primarily determined by the ratio of the OFF to ON switch resistances, wherefrom it is evident that $t_s/t_m >1$. The threshold parameters ($V_{th}$, $V_h$) may depend on the oscillation frequency, though indirectly, because of the heat balance variation, and this dependence is very weak. For the oscillation frequency used in our work, $t_s/t_m\sim(10-20)$; therefore, the semiconductor phase mainly contributes at EBM, and the metal phase – at recovery.

The results presented allow the use of EBM as a method for modifying the frequency of a single oscillator circuit, with the possibility of reversible recovery of the initial parameters by means of exposure to air. The main reason for the frequency change is the modification of parameters of the VO$_2$-switch ($V_{th}$, $R_{off}$). If we assemble a coupled network of such oscillators, EBM of switching structures can influence the dynamics of this system. Considering such circuits as prototypes of artificial neural networks, a mode of operation might be proposed, by analogy with [21], when, changing the parameters of the array of oscillators, they will operate as a system with pattern recognition, where the patterns are presented as a spatial distribution of the electron beam exposure dose. Thus, the results of this study not only show the physical aspects of electron beam modification of oscillators, but they demonstrate applied potentialities for the ONN implementation.

**Acknowledgements** This work was supported by Russian Science Foundation, grant no. 16-19-00135.